\documentclass[
    ,final            
  ,numberedheadings 
  ]
  {aipproc}

\layoutstyle{6x9}

\begin{document}

\title{Motion of the hot spot and spin torque in accreting millisecond pulsars}


\author{Alessandro Patruno}{
  address={Astronomical Institute ``Anton Pannekoek'', University of Amsterdam, Kruislaan 403,\\
    1098 SJ Amsterdam, the Netherlands; A.Patruno@uva.nl}
}

\begin{abstract}

The primary concern of this contribution is that accreting millisecond
pulsars (AMXPs) show a much larger amount of information than is
commonly believed.  The three questions to be addressed are: \\ 1. Is
the apparent spin torque observed in AMXPs real ? \\ 2. Why do we see
correlations and anti-correlations between fractional amplitudes and
timing residuals in some AMXPs ? \\ 3. Why the timing residuals, the
lightcurve and the 1Hz QPO in SAX J1808.4$-$3658 are related ? \\

\end{abstract}

\maketitle

\section{Motivation}

There has been a large debate in the accreting millisecond pulsar
community, on how to interpret the apparent measurement of spin
torques in accreting millisecond pulsars (AMXPs).  The need of
interpreting what the apparent spin torque really is needs hardly be
explained in this meeting, and this is also the reason why I call it
'apparent' spin torque. An improved understanding of the nature of the
apparent spin torque can have enormous implications to the knowledge
of neutron star physics.  For example the existence (or the lack) of a
spin torque has consequences in the gravitational wave
physics\citep{watts1} \citep{watts2}, on the problem of the break up
spin frequency of neutron stars\citep{Chakrabarty} and on the strength
of their magnetic fields. At the time of writing this contribution, we
know ten AMXPs, seven of which show persistent pulsations, and other
three are the so called intermittent AMXPs (see for example the
contribution of Altamirano \& Casella and Galloway for an
observational overview at this meeting, and
\citep{galloway}\citep{casella}
\citep{gavriil}\citep{altamirano}\citep{patruno}).  The primary target
of this contribution are the seven persistent AMXPs. In particular I
will discuss the apparent spin torque measurement in XTE J1751-305 and
XTE J1807-294 and show how a different and correct treatment of the
statistical errors can completely change the interpretation of the
observational results.  Related to the problem of the apparent spin
torque measurement is the interpretation of the relation between the
fractional amplitude of the pulsations and the residuals of the pulse
time of arrivals (TOAs).  The residuals here refer to the coherent
timing analysis of the AMXPs. In a few words: one measures the TOAs of
the pulsations and then fits a model (usually a keplerian circular
orbit plus a constant spin frequency or a spin torque) to the
TOAs. What is left after the fit are the residuals, and they
represent the relative location, with respect to an arbitrary
reference position, of the hot spot on the neutron star surface, i.e.,
the relative pulse phases. I will call this relation ``the FAR
relation'' throughout this contribution (FAR$=$Fractional Amplitude
vs. Residuals).  I will show that this relation exists in two sources,
XTE J1807-294 and XTE J1751-305, and that it is probably a more
general phenomenon among AMXPs.  But why the existence of the FAR
relation is connected with the interpretation of the apparent spin
torque ? The reason is quite simple: we measure the apparent spin
torque from the TOAs, and therefore whatever its origin is, it must be
related in some way with the fractional amplitudes of the pulsations.
Finally, I will discuss some intriguing relations between the
lightcurve, the periodic timing and the aperiodic timing of the AMXP
SAX J1808.4$-$3658. This AMXP shows some strange shifts of the pulse
phases in the timing residuals that cannot be explained even with a
torque model. Intriguingly, when the pulse phases start to shift the
slope of the lightcurve changes. And when the phases stop to shift,
the unexplained phenomenon of the 1 Hz QPO appears. The implications
of the relation between such different phenomena are profound: if the
unmodeled phase shifts are related with the lightcurve and the
aperiodic variability, then a deep link between the accretion disk
(responsible for the lightcurve and perhaps the QPO) and the surface
of the neutron star (pulsations) exists. Is it possible that all these
phenomena (apparent spin torque, FAR relation,
lightcurve-periodic-aperiodic variability link) have a common unified
explanation ? The search for an answer to this question is the
motivation of this work.

\section{The problem of the apparent spin torque}

The measure of a spin torque has been claimed in six out of seven
known AMXPs. The seventh AMXP has not a measured spin torque probably
because it has been observed only during the tail of its outburst.
Therefore the measure of the spin torque can appear a well established
and thorough result of AMXP physics.  My main concern however is that
the measure of the spin torque, at least in some AMXPs, is the result
of the combination of underestimated statistical errors plus a
possible misleading use of standard timing techniques.  The reason why
the statistical errors are usually underestimated when using standard
fitting techniques is a simple consequence of the so called
\emph{timing noise}.  The term \emph{timing noise} traces back to the
early '70s in the field of radio pulsar timing. In radio pulsar timing
one calculates the pulse TOAs and then fits a polynomial expansion
truncated at the second or third order. The first order (linear) term
of the polynomial is the spin frequency, the second order (quadratic
term) is the spin down due to dipole radiation, and, sometimes, the
third order term is related with the breaking index of the
pulsar. After removing these terms from the TOAs, one should expect a
gaussian distribution of the residuals with zero mean. The gaussian
distribution of the post-fit residuals should contain indeed only the
measurement error component due to the counting statistics.  However
in many young pulsars, the post-fit timing residuals had a very large
amount of quasi-sinusoidal structures that were indicating some
unmodeled component of unknown origin. All this unmodeled component is
what was called \emph{timing noise}. The term 'noise' can be a bit
misleading since it has nothing to do with the \emph{white noise}
expected from the measurement error component. It is called 'noise'
only because we do not have any satisfactory model to interpret its
presence in the post-fit residuals. Now with the AMXPs the story
repeats.  One fits a polynomial expansion truncated at the first or
second order (after removing the orbital component), and interprets
the linear term as the spin frequency, and the quadratic term as the
spin torque. However the post-fit residuals have not a gaussian
distribution, and lots of structures are particularly evident in some
of them. XTE J1807$-$294 is the AMXP that contains the largest amount
of timing noise, regardless if we consider a constant spin or a spin
torque model. On the other side, some AMXPs show very little timing
noise after removing a quadratic component from the TOAs. However
still some unmodeled structures are visible in the residuals and
cannot be overlooked, especially when calculating the significance of
an apparent spin torque measurement.  Indeed all the standard
$\chi^{2}$ minimization techniques used for the polynomial fit and the
statistical errors are based on least square numerical techniques, and
as such they use the hypothesis that the source of noise in white, or
in other words, that the distribution of the post-fit residuals is
gaussian with zero-mean.  However if there is also \emph{timing
noise}, the statistical errors obtained from the least square fit are
meaningless. A better way for estimating realistic statistical errors
is by means of Monte Carlo (MC) simulations, to take into account the
presence of correlated noise.  The concept behind the MC simulations
is very simple: we measure some spin frequency and a spin frequency
derivative (our candidate spin torque) plus some residual timing noise
that can be decomposed in a Fourier series and analyzed with standard
Fourier techniques.  The power spectrum we obtain is the 'fingerprint'
of the (unknown) physical process behind the TOAs. Now we generate
thousands of almost identical power spectra and transform them back
into timing residuals.  Since the power spectra contain no information
on the phases of the Fourier frequencies, we randomize the phases when
transforming back the power spectrum into fake timing residuals. We
then measure the spin frequency and its derivative in all the ensemble
of fake residuals and create a distribution of spin frequencies and
derivatives. The standard deviations of these two distributions are
our statistical errors on the spin frequency and on the spin frequency
derivative.  In other words we assume that all the variability we see
in the timing residuals is due to timing noise (including the
variability due to the quadratic polynomial term).  Then we try to
answer to the following question: given our assumptions, what is the
probability of measuring a spin frequency derivative identical to the
one we observe if we suppose that \emph{all} the variability we see is
due to timing noise ?  Of course this method has its limitations and
its assumptions, but nonetheless it takes into account the effect of
correlated (red) noise in the fit of the parameters. To show the
effect of MC simulations on the significance of the quadratic
component I will focus on two AMXPs: XTE J1807$-$294 (J1807 now on)
and XTE J1751$-$305 (J1751 now on).  The periodic and aperiodic
variability of the two AMXPs were studied in detail by
\citep{zhang06},\citep{linares},\citep{chou08},\citep{riggio07} and
\citep{riggio08} (for J1807) and \citep{papitto08}(for J1751).  The first one
is the best example of a noisy AMXP, while the latter is a low
noise AMXP.  In J1807 the pulse profiles have a significant 1st
overtone beside the fundamental frequency, so we analyzed the two
harmonics separately. We found a significant spin derivative for the
fundamental harmonic, even when taking into account the colored nature
of the noise in the timing residuals, but the significance of the spin
torque for the 1st overtone disappears.  However, as already discussed
in \citep{riggio08} and \citep{chou08}, the position error has to be
taken into account for this source, since the best available position
comes from {\it Chandra} observations whose $68\%$ confidence level
error circle is $0''.4$.  The systematic error introduced in this way
on the frequency and frequency derivative is respectively $3\times
10^{-8}\rm\, Hz$ and $\approx 0.7\times 10^{-14}\rm\, Hz/s$
(calculated with eq. A1 and A2 from \citep{hartman08}), that summed in
quadrature with the statistical errors give a significance of $\approx
2.7\sigma$ and $\approx 1.5\sigma$ when considering also the
astrometric errors for the two harmonics respectively.  According to
\citep{papitto08}, J1751 has a spin up measurement of
$\dot{\nu}=3.7(1.0)\times 10^{-13} \rm\,Hz/s$, where
the error has to be considered at the 90$\%$ confidence level.  If we
repeat the timing analysis for this source and we apply the MC method
just described, we obtain $\dot{\nu}=4.7(1.2)\times
10^{-13}\rm\,Hz/s$, where the error now is our $1\sigma$
uncertainty. We see that when considering the MC simulations, the
significance of the apparent spin torque has decreased, but it is
still above the $3\sigma$ acceptance level.  It is interesting to note
that without the MC simulations our statistical error on the fitted
parameter is $0.5\times 10^{-13}\rm\,Hz/s$, which is 2.4 times smaller
than our final $1\sigma$ uncertainty.  This shows how overlooking the
timing noise during the parameter fitting leads to unrealistic,
underestimated, parameter errors.  The existence of timing noise in
J1751 is reflected in the bad reduced $\chi^{2}$ of the fit when
using a spin torque model.  Indeed our reduced $\chi^{2}$ is close to
1.3, for 518 degrees of freedom. This means that we are
leaving a small, but significant, amount of unmodeled structures in
the residuals. If we try for example to fit a spin frequency second
derivative $\ddot{\nu}$, then, according to an F-test, we get a
significant measure of $\ddot{\nu}$.  The significance of $\ddot{\nu}$
means that it absorbs some residual variability still present in the
timing residuals.

\section{The FAR relation}

In the previous section we have shown that the quadratic term of the
fit (apparent spin torque) is significant in J1751, while it is not in
J1807 when considering the timing noise in the determination of the
statistical errors of the fit. Now we focus on the follow up of the
discussion: if the measured quadratic component is significant even
when considering the timing noise, what does it physically represent ?
A natural answer to this is the spin torque expected from the transfer
of angular momentum from the disk to the accreting neutron star.
However the observed correlations between pulse fractional amplitudes
and timing residuals throws a spanner in the works.  Let's first focus
on J1751 and let's apply the solution found by \citep{papitto08} but
without the spin torque (constant spin frequency model). Of course
this choice of a constant frequency model is somehow arbitrary,
because we are not re-fitting the spin frequency and therefore we are
not minimizing the rms of the residuals.  However with this choice we
are making the assumption that all the phase movements that we see in
the residuals are due to something unrelated with the spin
frequency. The phase wanderings in the phase residuals
have a similar shape as the fractional amplitude of the pulsations.
Equally interesting, the shapes of the residuals and the fractional
amplitudes are similar to the lightcurve.  In other words, the
residuals, the fractional amplitude and the X-ray flux show a linear
correlation.  It is impressive to see that this correlation rises from
the simple assumption that the phase wandering of the pulsations does
not reflect a spin frequency variation.  Moreover, while the shape of
the residuals depends on the choice of the spin frequency, the shape
of the fractional amplitudes and the X-ray flux is 'intrinsic' to the
process of emission, and cannot change according to the chosen timing
solution. In particular the pulse fractional amplitude shape (i.e.,
the shape of the F.A. curve vs. time) does not change if we fold the
pulsations with or without a spin torque.  The fact that the phase
residuals have the same shape of the lightcurve and of the fractional
amplitude is impressive but still, can be a simple coincidence or a
consequence of the arbitrary choices made in the data
analysis. Therefore let's look at J1807, to see if we observe the same
behavior. If we use a constant frequency model we observe a linear
anti-correlation in the FAR diagram of the fundamental, while nothing
is seen for the 1st overtone. So, differently from J1751, in J1807
the phase residuals of the fundamental are anti-correlated with the
fractional amplitudes. Here both the lightcurve, the fractional
amplitudes and the residuals have lots of structures so the existence
of the linear anti-correlation can hardy be justified with a simple
coincidence.  Also the lightcurve seems to be anti-correlated with the
phases, but again, with the assumption that the structures in the
residuals do not represent any variation of the spin frequency and the
rms of the residuals is not minimized during the fit.  Interestingly
the anti-correlation becomes tighter when removing a quadratic
component.  A similar \emph{anti-correlation} is observed in another
source: XTE J1814-338. This is also an AMXP with a large amount of
structures in the residuals (timing noise) and a strong 1st overtone
in the pulse profiles.  On the other side a similar
\emph{correlation}, as the one observed for J1751, exists in IGR
J00291+5934 if we use a constant spin frequency model: the phase
residuals, the fractional amplitude and the lightcurve follow a linear
relation. This is a source with a small amount of timing noise, no
significant 1st overtones in the pulse profiles, and a smooth
lightcurve, quite similar to J1751.
\begin{figure}[!hpb]
 \includegraphics[height=.3\textheight,width=1\textwidth, clip]{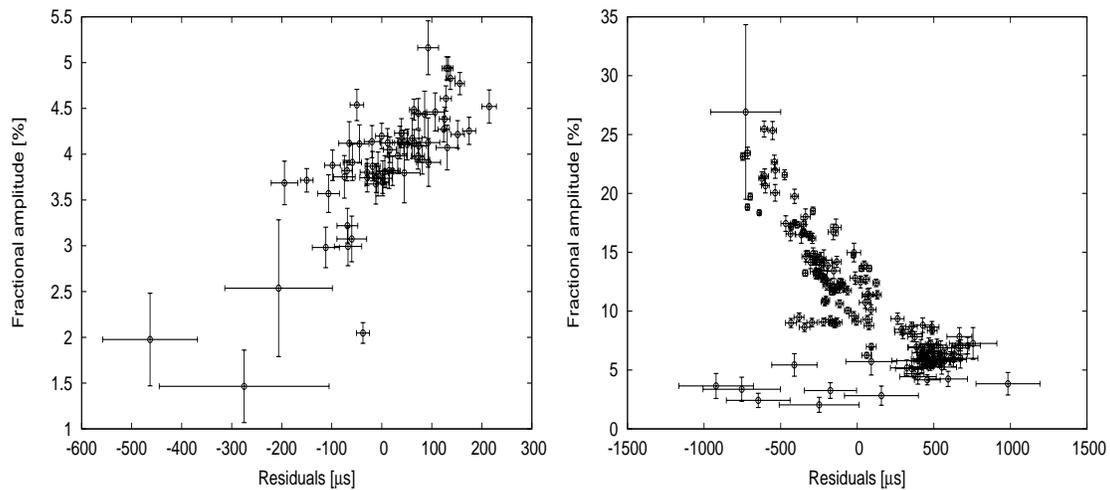}
  \caption{Fractional amplitude vs. Timing Residuals of two AMXPs: J1807 (right panel)
and J1751 (left panel). In the left panel there is a linear correlation between the fractional amplitudes 
and the residuals, while in the right panel we see an anti-correlation. The residuals of J1751 and J1807
were calculated assuming a constant frequency model. In the case of J1807 the anti-correlation becomes tighter 
if we remove also a quadratic term from the fit.}
\label{far}
\end{figure}

\section{The 1Hz QPO and the timing noise in SAX J1808.4$-3$658}

The reason why the FAR and the flux-residual relations can be an
essential ingredient in the AMXP physics is the tight link they imply
between the position of the emitting region (phase), the strength of
the pulsed signal received (fractional amplitude) and the accretion
disk physics (lightcurve). An outstanding example of this
interrelation is seen in the AMXP SAX J1808.4$-$3658 (J1808 from now
on). This source has shown five outbursts since the 1996, four of
which were observed with the {\it{RXTE}} satellite. During the 2000,
2002 and 2005 outbursts, J1808 has shown the puzzling phenomenon of
the 1Hz QPO.  This is a QPO with a fractional rms close to (or even
larger than) 100$\%$ that appears when the lightcurve reaches its
minimum and starts to bump in the so called re-flaring state. The
origin of this QPO is unknown, its fractional rms shows an energy
dependence, and it is observed always in the same position of the
lightcurve (re-flaring) in all three outbursts.  In the 1998 it was
not observed, probably because the re-flaring state of the lightcurve
was not observed with any X-ray telescope.  This QPO cannot originate
from the surface, as it is hard to explain its high rms, and also why
we do observe only one mode of oscillation.  It cannot be a QPO
originating from the disk alone, for example in a dip or shadowing of
the disk, since its energy dependence contradicts the expected flat
energy dependence.  
\begin{figure}[!hpb]
 \includegraphics[height=.3\textheight,width=1\textwidth, clip]{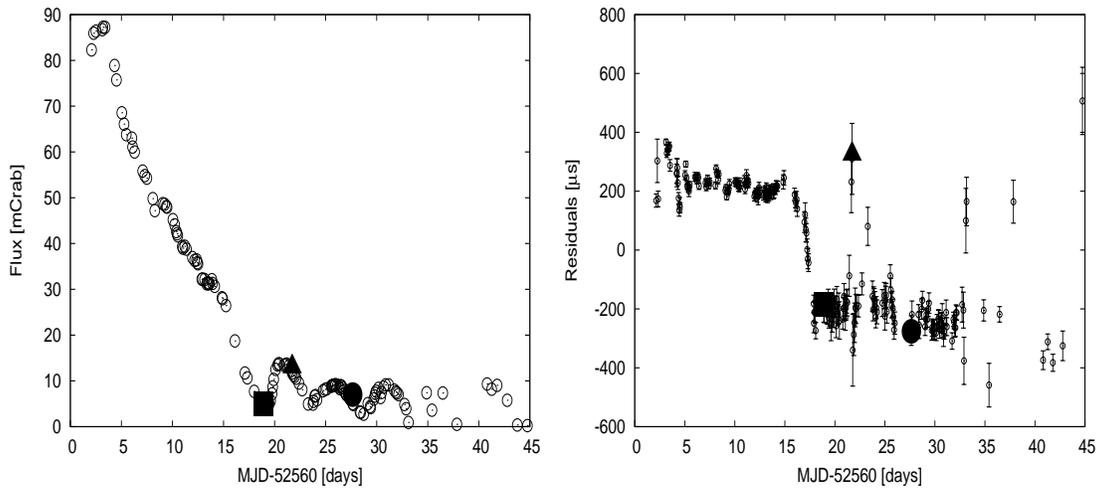}
  \caption{Left panel: lightcurve of the 2002 outburst of J1808. Right panel: timing
residuals (relative to a constant spin frequency model) of the 2002 outburst of J1808.
The solid square around MJD 52578 is the point where the pulse phases stop to drift
and the 1Hz QPO appears. The solid triangle at MJD 52582 is the point where the 1Hz QPO
disappears for the first time and the pulse phases jump by $\approx 0.3$ cycles. The solid 
circle at MJD 52587 is the point where the 1Hz QPO rms fractional amplitude reaches its maximum.
The behavior of the lightcurve, the QPO and the pulse phases is nearly identical in the 2005 outburst.}
\label{1Hz_2002}
\end{figure}
Probably its origin comes from a disk-magnetospheric interaction
associated with some kind of instability
\citep{aly}\citep{arons}\citep{spruit}\citep{elsner}\citep{romanova04}.
If we have a look at the pulse phases in the timing residuals,
obtained using a constant spin frequency model, then we see a phase
shift of approximately 0.2 cycles that was originally observed by
\citep{burderi06}.  This jump is not observed for the 1st overtone.
When the phases start to shift, the lightcurve changes slope and
decays faster before reaching the minimum flux.  At the minimum flux
the phases stop shifting and in coincidence the 1 Hz QPO appears. A
similar phase jump is also seen in the 2005 phase residuals, that
still show the same behavior. And the same
happens also for the decay point in the lightcurve and the appearance
of the 1Hz QPO. One peculiarity is that this QPO is not observed
persistently during the re-flaring state. It sometimes disappears, and
when this happens the pulsations jumps by $\approx0.3$ cycles, in both
the outbursts.  This is once more an indication that what is seen in
the lightcurve and in the aperiodic variability of the disk (QPO) has
an effect on the pulse phases.

\section{Conclusions}
We conclude that, when calculating a timing solution, all these effect
must be taken into account, otherwise we can confuse an effect of the
magnetospheric-disk interaction with a surface effect due to, for
example, a spin torque.  Of course the final question is how do we
take these effects into account when doing a coherent timing analysis.
An answer here cannot be satisfactory, since all the described
phenomena have not a clear theoretical explanation, yet. Certainly
they all show that the quadratic component usually identified with a
spin torque has not a higher dignity among the other terms of the
polynomial expansion used in standard coherent timing
techniques. Indeed it appears somehow arbitrary to treat the lowest
order variation (quadratic component) as a distinct entity with
respect to all the other phase shifts observed. This is particularly
evident after proving that the phases of J1808 are influenced by the
accretion disk-magnetospheric interaction. Another aspect to be
considered in AMXP physics is the possibility that the phase
variations we observe all come from the motion of the hot spot. There
are several simulations and models
\citep{romanova03}\citep{romanova04}\citep{romanova08},\citep{lamb08}
that predict a moving hot spot around a quasi-equilibrium position on
the neutron star surface. These movements can lead to an increase of
the fractional amplitude directly anti-correlated with the time of
arrival of the pulsations. However why the noisy AMXPs like J1807 and
J1814 show an anti-correlation while J1751 an J00291 do not, is not
easily explainable. Has the presence of a strong 1st overtone some
influence on the properties of the AMXPs ?  Another important point is
that, beside the theoretical model required to explain the
observations, one has to face a more serious problem that is how to
extract the correct informations from the data and very little can be
said until a satisfactory technique is developed.

\begin{theacknowledgments}
I'd like to thank A. Watts, J. Hartman, M. Klein-Wolt, Rudy Wijnands,
Michiel van der Klis and D. Chakrabarty, without whom this work would
not have been possible.  I thank also A. Papitto, and A. Riggio for
useful discussions we had during this enjoyable meeting.
\end{theacknowledgments}

\bibliographystyle{aipproc} 

\begin{thebibliography}{24}
\expandafter\ifx\csname natexlab\endcsname\relax\def\natexlab#1{#1}\fi
\providecommand{\enquote}[1]{``#1''}
\expandafter\ifx\csname url\endcsname\relax
  \def\url#1{\texttt{#1}}\fi
\expandafter\ifx\csname urlprefix\endcsname\relax\def\urlprefix{URL }\fi
\providecommand{\eprint}[2][]{\url{#2}}

\bibitem[{Watts} et~al.(2008)]{watts1}
A.~L. {Watts}, B.~{Krishnan}, L.~{Bildsten}, and B.~F. {Schutz}, \emph{\mnras}
  \textbf{389}, 839--868 (2008)

\bibitem[{Andersson} et~al.(2005)]{watts2}
N.~{Andersson}, K.~{Glampedakis}, B.~{Haskell}, and A.~L. {Watts},
  \emph{\mnras} \textbf{361}, 1153--1164 (2005),

\bibitem[{Chakrabarty} et~al.(2003)]{Chakrabarty}
D.~{Chakrabarty}, E.~H. {Morgan}, M.~P. {Muno}, D.~K. {Galloway},
  R.~{Wijnands}, M.~{van der Klis}, and C.~B. {Markwardt}, \emph{\nat}
  \textbf{424}, 42--44 (2003)

\bibitem[{Galloway} et~al.(2007)]{galloway}
D.~K. {Galloway}, E.~H. {Morgan}, M.~I. {Krauss}, P.~{Kaaret}, and
  D.~{Chakrabarty}, \emph{\apjl} \textbf{654}, L73--L76 (2007),

\bibitem[{Casella} et~al.(2008)]{casella}
P.~{Casella}, D.~{Altamirano}, A.~{Patruno}, R.~{Wijnands}, and M.~{van der
  Klis}, \emph{\apjl} \textbf{674}, L41--L44 (2008)

\bibitem[{Gavriil} et~al.(2007)]{gavriil}
F.~P. {Gavriil}, T.~E. {Strohmayer}, J.~H. {Swank}, and C.~B. {Markwardt},
  \emph{\apjl} \textbf{669}, L29--L32 (2007)

\bibitem[{Altamirano} et~al.(2008)]{altamirano}
D.~{Altamirano}, P.~{Casella}, A.~{Patruno}, R.~{Wijnands}, and M.~{van der
  Klis}, \emph{\apjl} \textbf{674}, L45--L48 (2008)

\bibitem[{Patruno} et~al.(2008)]{patruno}
A.~{Patruno}, D.~{Altamirano}, J.~W.~T. {Hessels}, P.~{Casella}, R.~{Wijnands},
  and M.~{van der Klis}, \emph{ArXiv e-prints} \textbf{801} (2008),
 
\bibitem[{Zhang} et~al.(2006)]{zhang06}
F.~{Zhang}, J.~{Qu}, C.~M. {Zhang}, W.~{Chen}, and T.~P. {Li}, \emph{\apj}
  \textbf{646}, 1116--1124 (2006)

\bibitem[{Linares} et~al.(2005)]{linares}
M.~{Linares}, M.~{van der Klis}, D.~{Altamirano}, and C.~B. {Markwardt},
  \emph{\apj} \textbf{634}, 1250--1260 (2005)

\bibitem[{Chou} et~al.(2008)]{chou08}
Y.~{Chou}, Y.~{Chung}, C.-P. {Hu}, and T.-C. {Yang}, \emph{\apj} \textbf{678},
  1316--1323 (2008)

\bibitem[{Riggio} et~al.(2007)]{riggio07}
A.~{Riggio}, T.~{di Salvo}, L.~{Burderi}, R.~{Iaria}, A.~{Papitto}, M.~T.
  {Menna}, and G.~{Lavagetto}, \emph{\mnras} \textbf{382}, 1751--1758 (2007),

\bibitem[{Riggio} et~al.(2008)]{riggio08}
A.~{Riggio}, T.~{Di Salvo}, L.~{Burderi}, M.~T. {Menna}, A.~{Papitto},
  R.~{Iaria}, and G.~{Lavagetto}, \emph{\apj} \textbf{678}, 1273--1278 (2008),

\bibitem[{Papitto} et~al.(2008)]{papitto08}
A.~{Papitto}, M.~T. {Menna}, L.~{Burderi}, T.~{di Salvo}, and A.~{Riggio},
  \emph{\mnras} \textbf{383}, 411--416 (2008)

\bibitem[{Hartman} et~al.(2008)]{hartman08}
J.~M. {Hartman}, A.~{Patruno}, D.~{Chakrabarty}, D.~L. {Kaplan}, C.~B.
  {Markwardt}, E.~H. {Morgan}, P.~S. {Ray}, M.~{van der Klis}, and
  R.~{Wijnands}, \emph{\apj} \textbf{675}, 1468--1486 (2008),

\bibitem[{Aly} and {Kuijpers}(1990)]{aly}
J.~J. {Aly}, and J.~{Kuijpers}, \emph{\aap} \textbf{227}, 473--482 (1990).

\bibitem[{Arons} and {Lea}(1976)]{arons}
J.~{Arons}, and S.~M. {Lea}, \emph{\apj} \textbf{210}, 792--804 (1976).

\bibitem[{Spruit} and {Taam}(1993)]{spruit}
H.~C. {Spruit}, and R.~E. {Taam}, \emph{\apj} \textbf{402}, 593--604 (1993).

\bibitem[{Elsner} and {Lamb}(1977)]{elsner}
R.~F. {Elsner}, and F.~K. {Lamb}, \emph{\apj} \textbf{215}, 897--913 (1977).

\bibitem[{Romanova} et~al.(2004)]{romanova04}
M.~M. {Romanova}, G.~V. {Ustyugova}, A.~V. {Koldoba}, and R.~V.~E. {Lovelace},
  \emph{\apj} \textbf{610}, 920--932 (2004)

\bibitem[{Burderi} et~al.(2006)]{burderi06}
L.~{Burderi}, T.~{Di Salvo}, M.~T. {Menna}, A.~{Riggio}, and A.~{Papitto},
  \emph{\apjl} \textbf{653}, L133--L136 (2006),

\bibitem[{Romanova} et~al.(2003)]{romanova03}
M.~M. {Romanova}, G.~V. {Ustyugova}, A.~V. {Koldoba}, J.~V. {Wick}, and
  R.~V.~E. {Lovelace}, \emph{\apj} \textbf{595}, 1009--1031 (2003),

\bibitem[{Romanova} et~al.(2008)]{romanova08}
M.~M. {Romanova}, A.~K. {Kulkarni}, and R.~V.~E. {Lovelace}, \emph{\apjl}
  \textbf{673}, L171--L174 (2008).

\bibitem[{Lamb} et~al.(2008)]{lamb08}
F.~K. {Lamb}, S.~{Boutloukos}, S.~{Van Wassenhove}, R.~T. {Chamberlain}, K.~H.
  {Lo}, A.~{Clare}, W.~{Yu}, and M.~C. {Miller}, \emph{ArXiv e-prints}
  \textbf{808} (2008)

\end{thebibliography}

\end{document}